\documentclass[preprint,12pt]{elsarticle}

\usepackage{longtable}
\usepackage{tabularx}
\usepackage{xcolor}
\usepackage[normalem]{ulem}
\usepackage{algorithm}
\usepackage{algorithmic}
\def\figureautorefname{Fig.}
\def\tableautorefname{Tab.}
\def\equationautorefname{Eq.}

\def\sectionautorefname{Sec.}
\def\etal{\textit{et al. }}
\usepackage{afterpage}
\usepackage{booktabs}
\usepackage{tabularray}
\usepackage{microtype}
\usepackage{multirow}
\usepackage{array}
\newcommand{\PreserveBackslash}[1]{\let\temp=\\#1\let\\=\temp}
\newcolumntype{C}[1]{>{\PreserveBackslash\centering}p{#1}}
\newcolumntype{R}[1]{>{\PreserveBackslash\raggedleft}p{#1}}
\newcolumntype{L}[1]{>{\PreserveBackslash\raggedright}p{#1}}
\usepackage{booktabs}

\usepackage{amssymb}
\usepackage{amsmath}

\journal{Journal of Computational Science}

\begin{document}

\begin{frontmatter}









\title{Precision Cancer Classification and Biomarker Identification from mRNA Gene Expression via Dimensionality Reduction and Explainable AI}


\author[1,+]{Farzana Tabassum}\ead{farzana@iut-dhaka.edu}
\author[1,+]{Sabrina Islam}\ead{sabrinaislam22@iut-dhaka.edu}
\author[1,3]{Siana Rizwan}\ead{sianarizwan@iut-dhaka.edu, siana.rizwan@queensu.ca}
\author[1,2]{Masrur Sobhan}\ead{msobh002@fiu.edu}
\author[1,3]{Tasnim Ahmed}\ead{tasnimahmed@iut-dhaka.edu,tasnim.ahmed@queensu.ca} 
\author[1]{Sabbir Ahmed}\ead{sabbirahmed@iut-dhaka.edu} 
\author[1,*]{Tareque Mohmud Chowdhury}\ead{tareque@iut-dhaka.edu}
\affiliation[1]{organization={Department of Computer Science and Engineering, Islamic University of Technology},
    addressline={Board Bazar}, 
    city={Gazipur},
    postcode={1704}, 
    country={Bangladesh}}
\affiliation[2]{organization={Knight Foundation School of Computing and Information Sciences (KFSCIS), Florida International University},
    city={Miami},
    postcode={FL 33199}, 
    state={Florida},
    country={USA}} 
\affiliation[3]{organization={School of Computing, Queen's University},
    city={Kingston},
    postcode={K7L 3N6}, 
    state={Ontario},
    country={Canada}}
\affiliation[*]{tareque@iut-dhaka.edu}

\affiliation[+]{(these authors contributed equally)}



\begin{abstract}
Gene expression analysis is a critical method for cancer classification, enabling precise diagnoses through the identification of unique molecular signatures associated with various tumors. Identifying cancer-specific genes from gene expression values enables a more tailored and personalized treatment approach. However, the high dimensionality of mRNA gene expression data poses challenges for analysis and data extraction. 
This research presents a comprehensive pipeline designed to accurately identify 33 distinct cancer types and their corresponding gene sets. 
It incorporates a combination of normalization and feature selection techniques to reduce dataset dimensionality effectively while ensuring high performance.  Notably, our pipeline successfully identifies a substantial number of cancer-specific genes using a reduced feature set of just 500, in contrast to using the full dataset comprising 19,238 features. By employing an ensemble approach that combines three top-performing classifiers, a classification accuracy of 96.61\% was achieved. Furthermore, we leverage Explainable AI to elucidate the biological significance of the identified cancer-specific genes, employing Differential Gene Expression (DGE) analysis.
\end{abstract}






\end{frontmatter}

\section{Introduction}

Cancer, a complex disease marked by the uncontrolled proliferation of cells \cite{wang2020regulatory, bekisz2020cancer}, remains a leading cause of mortality worldwide.  In 2022, it accounted for nearly 10 million deaths \cite{World_Health_Organization_2024}.  Gene expression analysis has emerged as a pivotal tool in this pursuit. Researchers have directed their efforts toward identifying potential biomarkers to tackle the challenges associated with cancer diagnosis and drug discovery. In response to the challenges associated with cancer diagnosis and drug discovery, researchers have focused on identifying potential biomarkers. encompassing the transcription of DNA into messenger RNA (mRNA) and the subsequent translation into proteins \cite{alharbi2023machine,yuan2023advances}. In recent years, Machine Learning (ML) methods have become increasingly popular for cancer cell classification due to their effectiveness in identifying important features and the availability of data from high-throughput machines \cite{kourou2015machine}.

In the domain of Cancer classification with RNA data, Podolsky \etal conducted an evaluation of four publicly available lung cancer datasets from various institutions, utilizing seven ML-based techniques \cite{podolsky2016evaluation}. Their findings indicated that performance varied significantly across the datasets, with the k-nearest neighbor (KNN) \cite{guo2003knn} and support vector machine (SVM) \cite{Wang2005SupportVM} methods achieving higher area under the curve (AUC) values \cite{huang2005using}. Additionally, experiments employing deep learning (DL) methodologies have been performed on mRNA datasets. Lyu \etal implemented a DL-based approach to classify tumor types using gene expression data derived from 33 tumors in the Pan-Cancer Atlas \cite{lyu2018deep}. Their classification model incorporated three convolutional layers with max-pooling and batch normalization \cite{ioffe2015batch}, followed by three fully connected layers.

Laplante et al. developed a deep neural network model to infer cancer locations using 27 miRNA data cohorts from The Cancer Genome Atlas (TCGA), categorized into 20 anatomical sites \cite{laplante2020predicting}. The authors employed log transformation and MinMaxScaler for data preprocessing, followed by a classification model built upon a six-layer artificial neural network (ANN) \cite{jain1996artificial}.
Hsu et al. utilized TCGA RNA-sequencing data to classify 33 distinct types of cancer \cite{hsu2018cancer}. Their methodology included feature selection techniques such as tree classifiers \cite{stein2005decision} and variance thresholding \cite{fida2021variance}, along with two data normalization approaches: min-max scaling and standard scaling. They showed a performance comparison using a decision tree, KNN, linear SVM (SVM) \cite{chang2008feature}, polynomial SVM, and ANN models where linear SVM outperformed other classifiers.
Mostavi et al. introduced three Convolutional Neural Network (CNN) models \cite{albawi2017understanding} for the classification of tumor versus non-tumor samples \cite{mostavi2020convolutional}. Their study indicated that the 1D-CNN model exhibited superior computational efficiency.
Mahin et al. proposed PanClassif, a method designed to enhance the performance of machine learning classifiers in cancer identification by utilizing a limited number of efficient genes extracted from RNA-seq data \cite{mahin2022panclassif}. Their approach incorporated six classifiers: SVM (linear kernel), SVM (RBF kernel), Random Forest \cite{breiman2001random}, Neural Network, k-Nearest Neighbor (KNN), and AdaBoost \cite{freund1999short}. Notably, none of these studies focused on the analysis and identification of significant gene sets or biomarkers specific to each cancer type within their datasets.

Focusing on the issue of identifying biomarkers for cancers, researchers conducted experiment with TCGA RNAseq data containing 33 types of cancer created a heatmap to represent gene scores, and used guided backpropagation and Grad-CAM \cite{selvaraju2017grad} to analyze genes \cite{de2019deepgx}. 
A gene functional classification method was utilized to evaluate 400 biomarkers and annotate genes by function similarity. The functional analysis results were compared to the cohort cancer types' relevant pathways using p-values for correlation with major gene pathways and biomarkers. 
In another study, Lopez-Rincon, A. \etal proposed an ensemble approach to extract 100 miRNA for multi-class cancer classification \cite{lopez2019automatic}. According to their findings, Logistic Regression \cite{huang2016feature} performed best across all of their experiments. They then conducted a bibliographical meta-analysis to confirm their findings. 

While several studies have aimed to identify cancer biomarkers, most have relied on ML and DL-based models as `black boxes' \cite{linardatos2020explainable}. Recently, various approaches have been developed to explain these black-box models, including Shapley Sampling \cite{vstrumbelj2014explaining}, Relevance Propagation \cite{bach2015pixel}, LIME \cite{ribeiro2016should}, ANCHOR\cite {ribeiro2018anchors}, DeepLIFT \cite{shrikumar2017learning}, etc. 
These explainability analysis techniques have widely been used in a wide spectrum of ML-based and DL-based tasks to enhance model transparency and trust, enabling researchers to leverage complex models while ensuring accountability and understanding \cite{ zhang2022discovering}.
In 2017, a game-theoretic approach known as SHAP (SHapley Additive ExPlanation) \cite{lundberg2017unified} was introduced to provide interpretability to black box models. Levy \etal utilized SHAP to identify significant methylation states across different cell types and cancer subtypes \cite{levy2020methylnet}.  Yap \etal utilized SHAP to provide explanations for a DL-based model that employed RNA-sequence data for cancer tissue classification \cite{yap2021verifying}. Divate \etal used SHAP to interpret the classification model and identify specific gene signatures that contributed to classifying different cancer types \cite{divate2022deep}. However, they only provided a certain number of features for the combination of all kinds of cancer.

Additionally, statistical validation in studying cancer-specific gene sets is essential to confirm their biological significance and eliminate the possibility of random chance. Differential Gene Expression (DGE) analysis is a frequently employed approach that identifies genes exhibiting significant expression level variations across various cancer types \cite{xue2020comprehensive,stupnikov2021robustness}. Ni \etal implemented DGEs to identify potential genes associated with the pathogenesis and prognosis of lung cancer, suggesting that DGEs can identify important genes across the samples \cite{ni2018identification}. 

To address the limitations of existing studies, which includes the lack of focus on identifying significant gene sets for specific cancer types, dimentionality curse of RNAseq data and the black-box nature different classifiers,  in this paper, we proposed an ensemble architecture to efficiently and accurately identify 33 different cancer types and their corresponding gene sets using mRNA gene expression data. We utilized a normalization and feature selection technique to enhance convergence efficiency and address the high dimensionality in gene expression data. Additionally, we implemented an interpretable machine learning tool to determine the significant global feature contribution responsible for the models' high performance and accuracy. The proposed approach demonstrates competence in achieving high accuracy with a limited number of 500 features only.

\section{Methods}\label{Methods}

Our study comprised two main portions: 1) Classification and 2) Identification of cancer-specific gene sets for each cancer type. In the classification stage, we addressed the challenge of high dimensionality in the dataset and aimed to reduce noise before applying classifiers. To achieve this, we explored various combinations of feature selection and normalization techniques. Additionally, we evaluated the performance of different standalone classifiers and implemented two ensemble techniques to enhance classification accuracy. 
To ensure a robust performance analysis for cancer classification, we employed Stratified 5-Fold cross-validation on the dataset. This approach was used to minimize bias and enhance the reliability of our results.
In the second stage of our study,  SHAP \cite{Lundberg2017AUA} was utilized to investigate the model explainability and identify significant genes specific to each cancer type. To assess the biological significance of our identified gene sets, we performed Differential Gene Expression (DGE) \cite{porcu2021differentially} analysis.

\begin{figure*}[t]
        \centering
        \includegraphics[width=\textwidth]{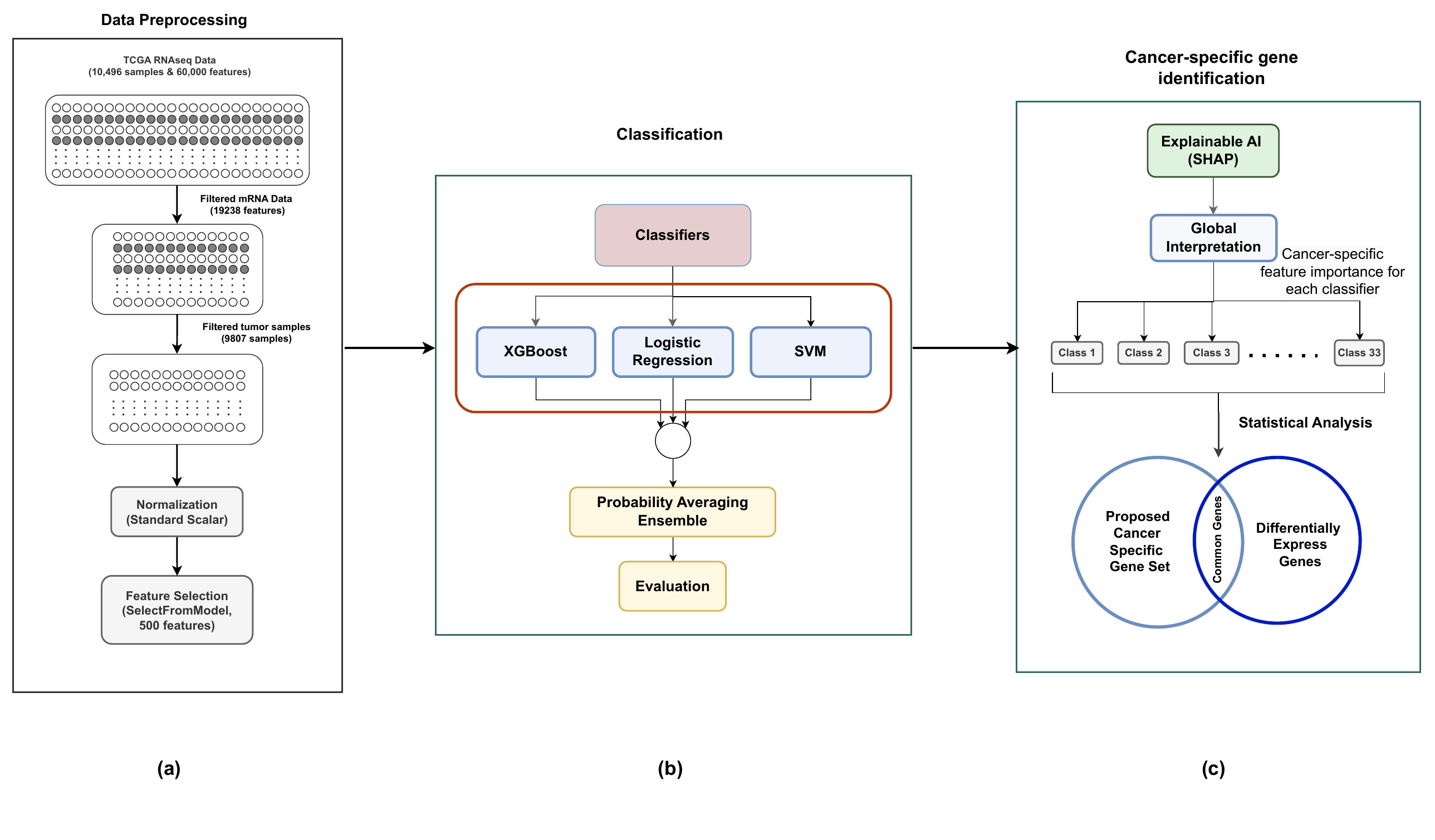}
\caption{Overview of the proposed pipeline. (a) Data preprocessing steps include selecting mRNA data from the TCGA dataset and reducing it to the top 500 mRNA genes using data normalization and feature selection. (b) Performed probability averaging ensemble technique by combining top 3 performed classifiers. (c) Analyzed model's performance using explainable AI (XAI) to identify cancer-specific gene sets and validated it through statistical validation.}
\label{pipelines}
\end{figure*}

After a thorough analysis, we proposed a final pipeline that yields the best results and a substantial number of biologically significant genes. The pipeline is visually represented in \figureautorefname~\ref{pipelines}.

 \subsection{Dataset}

We conducted the experiments on the widely used pan-cancer dataset which was downloaded from the TCGA (The Cancer Genome Atlas Program) Data Portal \cite{weinstein2013tcgar}. TCGA gene expression data generated using RNA-Seq is quantified using the FPKM (Fragments Per Kilobase of transcript per Million mapped reads) \cite{zhao2021tpm} metric and tools like TOIL (Transcriptome-Oriented Incremental Learning) and RSEM (RNA-Seq by Expectation Maximization) \cite{li2011rsem} are used to analyze and interpret this data. TOIL, RSEM, and FPKM data estimate the expression level of genes based on the sequencing reads of mRNA transcripts. In this study, we worked with RNA-seq FPKM data. There are a total of 10,496 patient samples along with their 60,499 genes. 
The Pareto principle \cite{bunkley2008joseph}, also known as the 80/20 rule or the principle of factor sparsity, was applied to divide the sample data into two sets: 80\% for training and 20\% for testing \cite{joseph2022optimal}. A validation set, which constitutes 20\% of the training set, is employed for the entire dataset. 
\tableautorefname~\ref{num_pf_samples} presents the total number of samples in each cancer type.

\footnotesize
\begin{longtable}[c]{clc}
\caption{Number of samples in each cancer type from the TCGA Dataset \cite{weinstein2013tcgar}}
\label{num_pf_samples}\\
\toprule
\textbf{\#} & \textbf{Cancer Type} & \textbf{Count}\\ \midrule
\endhead
\midrule
\endfoot
\endlastfoot
1 & Adrenocortical carcinoma (ACC) & 77 \\
2 & Bladder Urothelial Carcinoma (BLCA) & 426 \\
3 & Breast invasive carcinoma (BRCA)  & 1211 \\
4 & \begin{tabular}[c]{@{}l@{}}Cervical squamous cell carcinoma\\ endocervical adenocarcinoma (CESC)\end{tabular} & 309 \\
5 & Cholangiocarcinoma (CHOL)  & 45 \\
6 & Colon adenocarcinoma (COAD) & 329 \\
7 & Diffuse large B cell lymphoma (DLBCL)  & 47 \\
8 & Esophageal carcinoma (ESCA)  & 195 \\
9 & Glioblastoma multiforme (GBM) & 165 \\
10 & Head and Neck squamous cell (HNSC) & 564 \\
11 & Kidney Chromophobe (KICH)  & 91 \\
12 & Kidney renal clear cell carcinoma (KIRC)  & 603 \\
13 & Kidney renal papillary cell carcinoma (KIRP)  & 321 \\
14 & Acute myeloid leukemia (LAML) & 173 \\
15 & Brain Lower Grade Glioma (LGG) & 522 \\
16 & Liver hepatocellular carcinoma (LIHC)  & 421 \\
17 & Lung adenocarcinoma (LUAD) & 574 \\
18 & Lung squamous cell carcinoma (LUSC) & 548 \\  
19 & Mesothelioma (MESO) & 87 \\
20 & Ovarian serous cystadenocarcinoma (OV)  & 427 \\
21 & Pancreatic adenocarcinoma (PAAD) & 183 \\
22 & Pheochromocytoma and Paraganglioma (PCPG) & 185 \\
23 & Prostate adenocarcinoma (PRAD) & 548 \\
24 & Rectum adenocarcinoma (READ)  & 102 \\
25 & Sarcoma (SARC) & 264                        \\
26 & Skin Cutaneous Melanoma (SKCM) & 470 \\
27 & Stomach adenocarcinoma (STAD) & 450  \\
28 & Testicular Germ Cell Tumors (TGCT)  & 137 \\
29 & Thyroid carcinoma (THCA) & 571 \\
30 & Thymoma (THYM)  & 121 \\
31 & Uterine Corpus Endometrial Carcinoma (UCEC)  & 194 \\
32 & Uterine Carcinosarcoma (UCS) & 57 \\
 33 & Uveal Melanoma (UVM) & 79 \\  \midrule
 & \textbf{Total} & \textbf{10,496}            \\ \bottomrule
\end{longtable}

\normalsize

\subsection{Data Preprocessing}
This section explores the data preprocessing steps used for the classifications of 33 types cancers. This study  utilized the cancerous mRNA samples to capture the genetic profile of malignancies, normalization techniques to address differences in data magnitude, and feature selection methods aimed at reducing dimensionality and identifying the most relevant genes for cancer classification.
\subsubsection{Selecting Gene Identifiers and Excluding Normal Tissue Samples}
Initially, the dataset contained 10,496 patient tissue samples and each sample contained 60,499 gene expressions. In this study, the examination was restricted solely to mRNA characteristics. Following the exclusion of non-mRNA genes, the dataset comprised a total of 19,238 mRNA genes. Additionally, as one of the main objectives of this study was to identify cancer-specific gene sets, genes that exhibited predominant expression in tumor tissues while being absent or minimally expressed in normal tissue samples were considered. It was imperative to mitigate any potential impact that the existence of non-cancerous tissues may exert on the genetic profile of particular malignancies. The exclusion of normal tissue samples from the analysis reduced the likelihood of encountering false positive or negative results. After excluding all normal tissue samples, the total number of samples was 9807.

\subsubsection{Normalization}
When data exhibit significant differences in magnitudes, it can adversely affect the training of models. This can result in slower convergence or even complete failure of the model to converge. To address this issue, data normalization techniques can be employed \cite{henderi2021comparison, singh2020investigating, borkin2019impact}. Previous studies have utilized various approaches such as standard scaler, z-score, min-max scaler, and feature vector data normalization approaches \cite{polyakova2022data, raju2020study, swiderska2020impact}.
In our study, we specifically implemented two data normalization techniques: Min-Max Scaler \cite{panda2014smoothing, patro2015normalization} and Standard Scaler \cite{ahsan2021effect,bisong2019introduction,raju2020study} due to their superior performance \cite{chauhan2022performance, arafa2021regularized}. 

The min-max scaler involves scaling data to a predetermined range, often spanning from 0 to 1. It is susceptible to outliers and may lead to a reduction of data integrity in the tails of the data distribution.
Conversely, the standard scaler is a normalization technique that converts the data to possess a mean of zero and a variance of one. This normalization technique exhibits greater resilience in handling data with unknown ranges and is comparatively less susceptible to the influence of outliers. 
These techniques were applied as preprocessing steps to the dataset before performing classification tasks. Our objective was to assess the impact of these techniques on the performance of the classification models and choose the most suitable one.

\subsubsection{Feature Selection}
The large dimensionality of the data makes it essential to explicitly specify features while classifying cancer \cite{yesilkaya2022manifold}. Hence, in previous studies they have implemented different feature selection approaches to reduce data dimensionality by identifying pertinent genes that are linked to cancer and enhancing the efficacy of classification models as well \cite{lopez2019automatic,huljanah2019feature, fonti2017feature,liu2002comparative,hira2015review, yesilkaya2022manifold}.
In our study, we employed four distinct embedded feature selection methods \cite{brownlee2014introduction}, namely SelectKBest \cite{chauhan2020detection}, ElasticNet \cite{zou2003regression}, Lasso \cite{fonti2017feature, muthukrishnan2016lasso}, and SelectFromModel \cite{huljanah2019feature, agaal2022influence}, to ascertain the most pertinent features for the categorization of 33 diverse cancer types. It is imperative to clarify that the term ``feature" in this context pertains to the mRNA genes present in the dataset.

The SelectKBest method selects the top $k$ features using univariate statistical measures like chi-square or mutual information. The study used the $ANOVA~ F-value$\cite{kim2017understanding} scoring function for selection (\equationautorefname~\ref{eq:anova}). 
\begin{equation}
\mathrm{F} = \frac{\sum_{i} n_{i} \left(\overline{y_{i}} - \bar{y}\right)^{2}/ (k-1)}{\sum_{ij} \left(y_{ij} - \overline{y_{i}}\right)^{2} / (N-k)}
\label{eq:anova}
\end{equation}
\begin{align*}
\text{where, } \\
k &= \text{total number of groups}, \\
N &= \text{overall sample size}
\end{align*}

The ElasticNet (\equationautorefname~\ref{eq:elastic_net}) algorithm is popular in bioinformatics due to its ability to create parsimonious models with minimal non-zero weights \cite{friedman2010regularization}. It balances precision and weight magnitude through $L1$ and $L2$ regularization \cite{sokolov2016pathway,basu2018rwen}. In our experiment, we set the $L1$ ratio to 0.4 and selected top genes based on the number of features. 
\begin{equation}
\mathrm{F} = \frac{1}{m}\left[\sum_{l=1}^{m}\left(y^{(i)}-h\left(x^{(i)}\right)\right)^{2}+\lambda_{1} \sum_{j=1}^{n} w_{j}+\lambda_{2} \sum_{j=1}^{n} w_{j}^{2}\right]
\label{eq:elastic_net}
\end{equation}

\begin{align*}
\text{where, } \\
w_j & : \text{the weight for the } j^{\text{th}} \text{ feature,} \\
n & : \text{the number of features in the dataset,} \\
\lambda_1 & : \text{the regularization strength for the } L_1 \text{ norm,} \\
\lambda_2 & : \text{the regularization strength for the } L_2 \text{ norm.}
\end{align*}

The LASSO (\equationautorefname~\ref{eq:lasso_regularization}) technique is a regression analysis approach that incorporates variable selection and regularization to enhance the predictive accuracy and interpretability of the resulting statistical model \cite{tibshirani1996regression}. We utilized an $alpha$ value of 0.5 to identify the specified number of genes.
\begin{equation}
\text{Objective Function} = \text{Mean Squared Error} + \alpha \sum_{i} \left|w_i\right|
\label{eq:lasso_regularization}
\end{equation}
\begin{align*}
\text{where, } \\
\alpha &: \text{the hyperparameter controlling the strength of the L1 regularization term,} \\
w_i &: \text{weights (coefficients) assigned to each feature.}
\end{align*}

The SelectFromModel feature selection approach relies on a specified estimator to identify the most significant features. This approach has the advantage of autonomously identifying significant features without requiring extensive prior expertise. The function receives an estimator and a pre-defined limit on the number of features to be selected.

The techniques were applied to the dataset that underwent standard scaler normalization to determine the top 100, 250, 500, 750, and 1000 genes that are the most significant to identify each cancer type. Subsequently, the chosen genes were employed in a classification framework to categorize the 33 distinct cancer types present in the dataset.

\subsection{Standalone Classifiers}
In the case of cancer classification, the choice of appropriate classifiers can reduce the dimensionality curse and achieve higher accuracy \cite{mahin2022panclassif, de2019deepgx,  khalsan2022survey, linares2021machine, tufail2021deep, desdhanty2021liver, silva2020pan}. In order to identify the optimal classifiers for our dataset in cancer classification, we explored seven distinct methodologies: Random Forest \cite{breiman2001random}, XGBoost \cite{chen2016xgboost}, Logistic Regression \cite{huang2016feature}, SVM \cite{Wang2005SupportVM}, MLP \cite{svozil1997introduction}, 1-D CNN \cite{srinivasamurthy2018understanding}, and TabNet \cite{arik2021tabnet}. This approach allowed us to investigate different sectors, including traditional machine learning approaches, decision tree-based approaches, deep learning, and transfer learning. By doing so, we aimed to determine which type of classifier performs best on our dataset. To maintain consistency in the comparison, the parameters used for model fitting were held constant throughout all experiments.

\textbf{Logistic Regression: }
Logistic regression estimates class probabilities by fitting binary regression models and learned feature weights, predicting the class with the highest probability. In our logistic regression model, we implemented 100 maximum iterations to optimize the loss function and achieve convergence and $L2$ or ridge regularization \cite{hastie2020ridge} to handle outliers and model complexity. Additionally, to avoid overfitting, regularization \cite{bach2008bolasso} strength is set to 100.

\textbf{Support Vector Machines (SVM): }
SVM is applied to accurately categorize gene expression data by determining an optimal hyperplane with a maximum margin and identifying support vectors closest to the decision boundary \cite{guyon2002gene,alba2007gene}. In our study, a tolerance of $1e-5$ was set to ensure precise convergence during the optimization process. The ``one-vs-one" \cite{galar2011overview} decision function was implemented to create binary classifiers for each class pair and use a voting technique for final predictions.

\textbf{XGBoost: }
XGBoost exhibits exceptional performance in structured tabular data classification by leveraging a gradient-boosting \cite{friedman2002stochastic} algorithm. It optimizes performance through parallel processing, tree pruning, handling missing values, and regularization, effectively mitigating bias and overfitting. For our experimental setup, we limited the maximum depth of each decision tree to four. A learning rate of 0.1 was used to control the step size during the optimization process. Additionally, we configured the number of estimators to be 1000. When the difference between accuracy and loss remained consistent for 10 consecutive epochs, we stopped the training procedure.
These values were employed to minimize overfitting, reduce training time, and improve accuracy in the model.

\textbf{Random Forest: }
The Random Forest algorithm constructs multiple decision trees by randomly selecting subsets of features and samples. 
 The number of estimators was set to 20 in our classification step  to achieve optimal accuracy while avoiding overfitting the model.

\textbf{Multilayer Perceptron (MLP): }
MLP is a feedforward \cite{serre2007feedforward} neural network with input, hidden, and output layers. In gene expression analysis, the input is gene expression profiles, and the output layer predicts class probabilities \cite{ravindran2023survey,guillen2016cancer,gao2019deepcc}. In our study, we applied three hidden layers with 100 neurons each. An $alpha$ value of 0.001 was used to balance the capturing of data patterns and prevent overfitting. A learning rate of 0.001 ensured gradual convergence. The Rectified Linear Unit (ReLU) \cite{xu2015empirical} activation function captured non-linear associations.  Weight optimization was performed using the `$Adam$' optimizer \cite{kingma2014adam}. These hyper-parameter choices aimed to balance model complexity and generalization to new data.

\textbf{1D-CNN (1-Dimensional Convolutional Neural Network): }
For predicting cancer types, various convolutional neural network (CNN) models have been proposed \cite{mostavi2020convolutional,mohammed2021stacking}. 
Through experimentation, we found that using three fully connected layers with 512, 256, and 128 nodes respectively helped us achieve global minima in terms of loss. Batch Normalization was added between layers to standardize input and stabilize learning, reducing training epochs for faster convergence. The ReLU activation function was used to improve optimization and generalization. A dropout layer between the two layers worked as a regularizer to avoid overfitting. To 
 downsample data average pooling was applied. The final output layer of the model was a densely connected layer with a softmax activation function \cite{Goodfellow2018DeepLD} and had 33 nodes representing the class labels. Sparse categorical cross-entropy loss, Adam optimizer with a learning rate of 0.001, and accuracy as the evaluation metric were employed.

\textbf{Attentive Interpretable Tabular Learning neural network (TabNet): }
The effectiveness of the transformer architecture for cancer classification \cite{gokhale2023genevit,khan2021gene,zhang2022transformer} is due to its ability to capture long-range dependencies and contextual relationships from gene expression data. For this experiment, we used the TabNet classifier which incorporates an attention mechanism for working with tabular data \cite{rahman2022detection, mclaughlin2023fast, nasimian2023deep}. Its dynamic feature selection process enhances interpretability which improves generalization and reduces overfitting. In our study, the model used the $Adam$ optimizer along with a decay rate ($gamma$) of 0.9. $StepLR$ scheduler with a multiplication factor of 10 was employed to adjust the learning rate during training. Early stopping was implemented with a training threshold of 150 epochs. Batch size and virtual batch size were set to 512, and equal weighting was given to all training instances.

\subsection{Ensemble Approach}
An ensemble method is predicated on the notion that collective decisions are frequently superior to individual decisions \cite{xiao2018deep,tan2003ensemble}. Additionally, by reducing the impact of outliers or noisy data points, the use of ensemble techniques has the potential to improve stability and dependability. In this study, we implemented two types of ensemble approaches— Max-Voting \cite{assiri2020breast} and Probability-Averaging \cite{hosni2019reviewing}.
In the case of max voting, each model generates a prediction and the class label with the highest number of votes is chosen as the ultimate prediction. 
It is utilized to obtain a more precise estimation of class probabilities, rather than relying on the majority vote. This methodology incorporates model confidence and assigns greater weight to dependable predictions, leading to improved accuracy and calibrated probability estimation.

We selected an odd quantity of classification models to employ ensemble techniques. An odd number of models in an ensemble guarantees a majority class when voting based on their predicted class labels. This can facilitate decision-making and mitigate the occurrence of ties, which may arise when an even number of models are evaluated. Also, ensembling an odd number of models can effectively ignore outlier predictions from a single model and rely on the majority of predictions. 

\subsection{Cancer-specific gene set identification }
Understanding the significant features of each cancer type is crucial for unraveling molecular mechanisms, identifying novel biomarkers, and discovering potential therapeutic targets. Explainable AI methods offer a powerful approach for extracting feature importance and understanding the contribution of individual genes in the prediction of cancer samples \cite{lundberg2017unified}. In this study, we utilized SHAP \cite{Lundberg2017AUA} as the explainable AI method. SHAP is based on game theory principles and offers a mathematical approach to elucidate machine learning model predictions by calculating the individual impact of each feature. To calculate feature importance using SHAP, we passed the test data to the explainer along with the trained model and the explainer produced SHAP values, indicating the impact of each feature on the output prediction for each instance in the test data.

\subsubsection{Identifying the correctly predicted samples}
In the analysis of feature importance using SHAP scores from a model, we focused exclusively on the values from the correctly predicted samples. This is necessary because the SHAP scores of features from correctly predicted samples can provide genuine insights into the importance of each feature for a specific cancer. Thus, the genes that consistently and positively influence the model's ability to make accurate predictions for each cancer type can be prioritized \cite{futagami2021pairwise}. Including SHAP scores from incorrectly predicted samples can introduce misleading information, as those scores may not be relevant to that cancer and can lead to misinterpretation of the gene contributions. By considering only the SHAP scores from correctly predicted samples, we ensured more accurate and reliable identification of the significant features that are more likely to be biologically relevant and specific to each cancer type \cite{yap2021verifying}.

\subsubsection{Global interpretation of features using SHAP:} \label{Global interpretation}
SHAP provides feature attribution by calculating scores of each feature of all samples across various types of cancers. To focus on a specific cancer type, we identified the samples belonging to that particular cancer and considered only the SHAP scores of those samples for analysis. In order to evaluate the global importance of each feature for a specific cancer, we computed the median value for each feature individually across the identified samples specific to that particular cancer type. This process was repeated for all 33 types of cancers, enabling us to determine the median score for each feature across the samples associated with each cancer type. This approach allowed us to precisely assess the significance of each feature in the context of different cancer types and identify cancer-specific significant genes.

We utilized specific SHAP explainers tailored to each algorithm to calculate the SHAP scores of features for models trained by different classifiers. The LinearExplainer from SHAP was employed for Logistic Regression and Support Vector Machine (SVM) models. For XGBoost and Random Forest models, we utilized the TreeExplainer. Lastly, the DeepExplainer was used for Multilayer Perceptron (MLP) and 1D Convolutional Neural Network (1D-CNN) models.



\subsection{Statistical Validation using Differential Gene Expression}

In addition to employing different approaches for cancer classification, researchers utilized statistical tools such as DESeq2 \cite{porcu2021differentially}, edgeR \cite{sun2020differential}, LIMMA \cite{shriwash2019identification}, etc. to identify Differential Gene Expression (DGE).  
Differential gene expression (DGE) analysis facilitates the detection of genes with significant expression level variations across different cancer types \cite{stupnikov2021robustness,mcdermaid2019interpretation}. Sobhan \etal and Hossain \etal implemented DESeq2 for statistical genomic analysis and quantification of differential gene expression \cite{sobhan2022explainable,hossain2021pan,xue2020comprehensive}. 
DESeq2 is a practical and widely used tool for analyzing differential gene expression in RNA-seq data. It employs negative binomial generalized linear models and applies empirical Bayes approaches to estimate priors for log fold change and dispersion, providing posterior estimates for these values \cite{love2014moderated}.

The differential expression analysis with DESeq2 involves several steps. It utilizes normalization factors (size factors) to model raw counts and estimates gene-wise dispersion, enhancing the accuracy of dispersion estimates for count simulation. 
For our analysis, we utilized raw counts of gene expression values from both tumor samples and healthy tissue samples \cite{wang2017enabling,wang2018unifying}. Low-count genes were removed, retaining rows with at least 10 reads. The factor level was set to ``healthy tissue''. Our analysis was performed on individual samples of each cancer type to identify cancer-specific genes with a significant impact on each type of cancer. These genes capture the overall behavior of the population concerning each cancer type and can be considered as global features. We set the threshold for differential expression as $|log2Fold-change| > 3$ and an adjusted $p-value < 0.001$, ensuring a stringent selection of genes exhibiting substantial expression changes that are statistically significant. \tableautorefname~\ref{deseq2 count} depicts a comprehensive summary of the sample sizes for both tumor and healthy tissues, along with the aggregate count of genes that exhibit differential expression across 17 distinct cancer types.

\begin{table}[t]
\centering
\caption{Total number of differentially expressed genes using DESeq2 for each cancer class}
\label{deseq2 count}
\footnotesize
\begin{tabular}{C{0.5cm} C{1.5
cm} C{2.5cm} C{3cm} C{3cm}}
\toprule
\textbf{\#}  & \textbf{Cancer type} & \textbf{Tumor sample count} & \textbf{Healthy tissue sample count} & \textbf{Total differentially expressed gene} \\ \midrule
1  & BLCA & 364 & 19 & 802 \\
2  & BRCA & 984 & 112 & 707 \\
3  & CHOL & 33 & 11 & 1904 \\
4  & COAD & 287 & 43 & 878 \\
5  & ESCA & 185 & 13 & 818 \\
6  & HNSC & 462 & 44 & 889 \\
7  & KICH & 62 & 27 & 1449 \\
8  & KIRC & 477 & 74 & 821 \\
9  & KIRP & 238 & 31 & 768 \\
10 & LIHC & 297 & 50 & 737 \\
11 & LUAD & 505 & 61 & 1026 \\
12 & LUSC & 491 & 53 & 1769 \\
13 & PRAD & 428 & 50 & 266 \\
14 & READ & 89  & 12 & 883 \\
15 & STAD & 382 & 35 & 637 \\
16 & THCA & 443 & 55 & 445 \\
17 & UCEC & 143 & 25 & 1205 \\ \bottomrule
\end{tabular}
\end{table}
\normalsize

\section{Results}\label{sec2}
\subsection{Experimental Setup}

Both Python and R programming languages were employed at different stages of the study. Python facilitated the majority of the experimental processes, while RStudio was utilized specifically to apply DESeq2, a widely used R package, for the identification of differentially expressed genes (DEGs) in individual cancer types. All experiments were conducted on a system equipped with an Intel® Core™ i9-12900K CPU and dual NVIDIA GeForce GPUs, each with 24GB of memory.

For evaluating our classification models and proposed pipeline, we employed four standard metrics: Accuracy, Precision, Recall, and F1-score. These metrics provided a comprehensive evaluation of model effectiveness.

Accuracy quantifies the proportion of correct predictions relative to the total number of predictions made by the model. However, in the context of imbalanced datasets, this metric may yield misleading results. Precision measures the model's ability to correctly identify positive instances, while Recall evaluates the model's capacity to detect true positive instances among all actual positive cases. It is important to recognize that these evaluation metrics may not fully capture model performance in imbalanced datasets, where one class may significantly dominate the others. Such metrics can exhibit bias toward the majority class, potentially resulting in deceptive conclusions.

Accuracy measures the ratio of correct predictions made by a model to the total number of predictions. However, imbalanced datasets can lead to misleading results. Precision refers to the model's ability to accurately identify positive instances. Recall assesses the model's capability to accurately detect positive instances among all the true positive instances. 
It is important to note that these evaluation metrics may not provide a complete assessment of the model's performance in the case of an imbalanced dataset, where one class is much more dominant than the other. These metrics are biased toward the majority class, which can result in misleading outcomes. The F1 score is a more informative metric for assessing model performance in such situations. It is important in cancer classification as it considers false positives and false negatives, which are crucial in medical diagnosis. The F1 score is more helpful in the case of imbalanced datasets where the minority class is of interest. It takes into account both recall and precision to offer a balanced assessment of a model's performance on imbalanced datasets by weighing the trade-off between two metrics.


\subsection{Performance on the Baseline Dataset}
The dataset consisting of all the 19,238 mRNA features and 9807 tumor samples was considered as the baseline dataset. Initially, we evaluated the performance of the seven classifiers on this baseline dataset. The results are summarized in \tableautorefname~\ref{Performance evaluation table}. 

Among all the classifiers, Logistic Regression achieved the highest accuracy of 96.43\% and an F1 score of 0.9391. Additionally, SVM and XGBoost also demonstrated strong performances, surpassing 95\% accuracy  with satisfactory F1 scores. It is worth noting that although the DL models (MLP and 1D CNN) achieved high accuracies (95.34\% and 94.50\% respectively), their F1 scores were not as impressive. It indicates that while the model is good at making overall predictions, it struggles to account for class disparities and misclassifies certain minority classes.
\begin{table}[t]
\centering
\caption{Performance analysis of the standalone classifiers and the two ensemble techniques on the dataset containing 19238 features and the performance of our proposed pipeline (implementing Standard Scaler normalization and SelectFromModel feature selection techniques and Probability Averaging Ensemble combining Logistic Regression, SVM and XGBoost) on 500 selected features. }
\label{Performance evaluation table}
\footnotesize
\begin{tabular}{C{2.2cm} L{7cm} C{1cm} C{1cm}}
    \toprule
    \textbf{Number of Features} &
    \textbf{Classifier Models} &
    \textbf{Acc (\%)} &
    \textbf{F1-Score}\\

     \midrule
    \multirow{7}{*}{19238} 
    & Random Forest & 92.17 & 0.9033 \\
    & 1D-CNN & 94.5 & 0.844 \\
    & Tabnet & 94.56 & 0.9129 \\
    & MLP & 95.34 & 0.8673 \\
    & XGBoost & 95.52 & 0.9203 \\
    & SVM & 96.3 & 0.9387 \\
    & Logistic Regression & 96.43 & 0.9391 \\ \cline{2-4} 
    
    & Max Voting Ensemble (Logistic Regression, SVM, XGBoost)
    & 96.33 & 0.9386 \\  
    
    & Probability Averaging Ensemble (Logistic Regression, SVM, XGBoost) & \textbf{96.45} & \textbf{0.9399} \\ \midrule
        
    \multirow{4}{*}{500}     
    & XGBoost               &   95.10   & 0.9315 \\ 
    & SVM                   &   95.87   & 0.9344 \\
    & Logistic Regression   &   96.31   & 0.9382 \\
    \cline{2-4}  
    & Ours (using probability averaging ensemble) & \textbf{96.61} & \textbf{0.9415} \\ \midrule 
    
    \end{tabular}
\end{table}
\normalsize

To further enhance the overall performance, we adopted Probability averaging and Max-Voting ensemble approaches. For the ensemble method, we combined the predictions of the top three performing models, namely Logistic Regression, SVM, and XGBoost. These three models were chosen as they all achieved accuracies above 95\% with good F1 scores. After implementing the ensemble approaches, Probability averaging outperformed Max-Voting, resulting in a slight improvement in accuracy to 96.45\%. It also surpassed the individual accuracy of the Logistic Regression model. Hence, the highest accuracy achieved on the baseline dataset was 96.45\%.

We conducted a comprehensive evaluation using feature selection methods in order to tackle the high dimensionality issue posed by the dataset's 19,238 mRNA features. Our pipeline involved applying standard scaler normalization technique followed by the SelectFromModel feature selection technique to identify the most informative 500 features out of the original 19,238. 

Remarkably, despite the significant reduction in feature space, Logistic Regression displayed great performance, achieving an impressive accuracy of 96.31\%. 
This accuracy level not only outperformed the other classifiers but also came very close to the accuracy   obtained on the baseline dataset with all 19,238 features. SVM and XGBoost classifiers also managed to achieve accuracies higher than 95\% with the reduced set of features. It is important to highlight that the accuracy values were averaged across 5 folds, ensuring the consistency of the results.
Additionally, we utilized ensemble techniques by combining the predictions from the top-performing classifiers to enhance the overall performance. The results showed that Probability Averaging led the way, achieving the highest accuracy of 96.61\%. F1-score of 0.9415 indicates a robust and balanced performance of the model in correctly identifying all the different cancer classes. Notably, this accuracy even surpassed the highest accuracy attained on the baseline dataset. 

A particularly impressive aspect of this achievement is the drastic reduction in computational expenses. By utilizing selective features obtained through the feature selection technique, we were able to streamline the computational burden without compromising on accuracy.


\normalsize
\subsection{Ablation study}
The results of all the experiments conducted were meticulously analyzed to determine the optimal combination of normalization techniques, feature selection methods, and classifier models that yield the best performance. In this analysis, the accuracies and the F1 scores of the experiments were carefully examined. 
\begin{table}[ht]
    \centering
    \caption{Performance analysis of the classifiers (in terms of Accuracy) after applying Standard Scaler and MinMax Scaler normalization techniques using reduced feature set}
    \label{normalization results}
    \footnotesize
    \begin{tabular}{L{3cm} C{3cm} C{3cm}}
        \toprule
        \multirow{2}{*}{\textbf{Classifier}} & \multicolumn{2}{c}{Normalization Technique}\\ \cmidrule(l){2-3} 
        & \textbf{Standard Scaler} & \textbf{MinMax Scaler} \\ \midrule
 
        Logistic Regression & \textbf{96.31} & 96.17 \\
        SVM & \textbf{96.03} & 96.02\\
        XGBoost & \textbf{95.52}    & 95.47 \\
        MLP & \textbf{94.79}        & 84.46 \\
        1D-CNN & 94.6               & \textbf{94.96}\\
        Random Forest & 91.65       & \textbf{92.23}\\ 

        \bottomrule
    \end{tabular}
\end{table}

\subsubsection{Choice of Normalization technique}
After obtaining the performance on the baseline dataset, we applied different techniques to address the high dimensionality issue while maintaining the accuracy achieved with all the features. An ablation study was conducted to assess the impact of various normalization techniques, feature selection methods, and classifier models on overall performance. We evaluated various combinations to identify the most effective combination of these techniques for achieving the best performance with minimal features. At first, we applied the two normalization techniques. The results are shown in \tableautorefname~\ref{normalization results}. We observed that Standard Scalar enhances the performance better than MinMax Scaler for 4 out of the 6 classifiers. So we have chosen Standard Scalar as our normalization technique and proceeded for further implementations.

        

\normalsize

\subsubsection{Choice of Feature Selection technique}
After applying the normalization technique, we have implemented four different feature selection techniques with 1000 features. We chose 1000 feature count as a starter as working with around 5\% features reduces the computational cost greatly. The results after applying different feature selection methods of the three top  performing models are shown in \tableautorefname~\ref{Feature selection results}. From the results, it is observed that SelectFromModel feature selection methods performs the best among the four feature selection methods. For this reason, we chose SelectFromModel as our feature selection method.
\begin{table}[ht]
    \centering
    \caption{Performance analysis of the three best-performing classifiers with 1000 features chosen by different feature selection methods using `Standard Scaler' feature normalization technique.}
    \label{Feature selection results}
    \footnotesize
    \begin{tabular}{L{3.2cm} L{4cm} C{1cm}}
    \toprule
    \textbf{Feature Selection Technique} & \textbf{Classifier} & \textbf{Acc(\%)} \\ \midrule
    \multirow{3}{*}{SelectFromModel} & Logistic Regression & \textbf{96.49} \\
    & SVM & \textbf{95.87} \\ 
    & XGBoost & \textbf{95.10} \\ \midrule

    \multirow{3}{*}{Select-K-Best}  & Logistic Regression & 95.51 \\ 
    & SVM & 95.14 \\ 
    & XGBoost & 94.16 \\ \midrule

    \multirow{3}{*}{Lasso}& Logistic Regression & 95.47 \\
    & SVM & 95.38 \\
    & XGBoost & 94.87 \\ \midrule
    
    \multirow{3}{*}{ElasticNet}& Logistic Regression & 94.28\\
    & SVM & 94.29 \\
    & XGBoost & 93.58 \\ \bottomrule
    
    \end{tabular}
\end{table}

\subsubsection{Selecting the appropriate number of features}

\begin{figure*}[t]
        \centering
        \includegraphics[width=\textwidth]{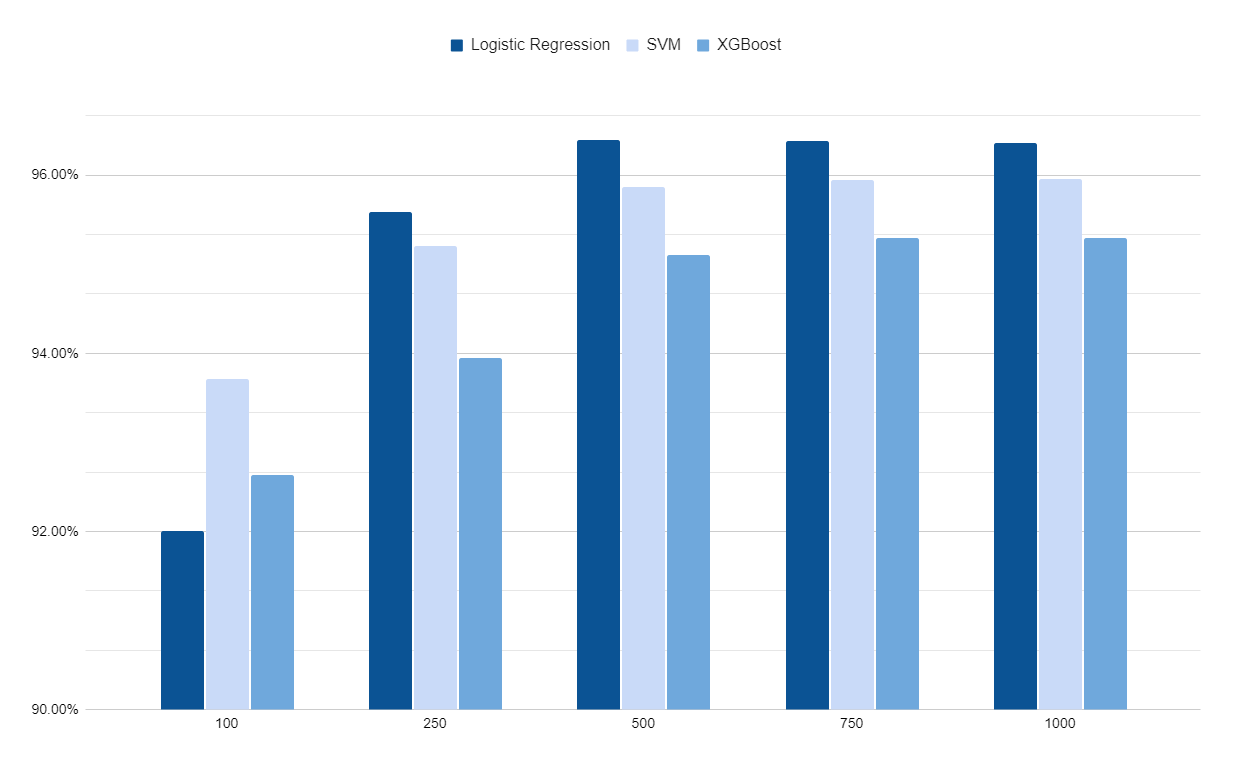}
\caption{The performance of the three best-performing classifiers with 100, 250, 500, 750 and 1000 features chosen by `SelectFromModel' feature selection method using `Standard Scaler' feature normalization}
    \label{feature count table}
\end{figure*}

\normalsize

Experiments were performed to find out how different feature selection affects the overall performance. Using the feature selection approaches, it was observed that using 100 and 250 features did not yield satisfactory outcomes, while performance significantly improved with 500, 750, and 1000 features. Interestingly, the top three models (Logistic regression, SVM, XGBoost) exhibited nearly identical accuracies across the 500, 750, and 1000 feature sets (\figureautorefname~\ref{feature count table}). Consequently, the decision was made to proceed with 500 features, striking a balance between computational efficiency and satisfactory performance.

\normalsize

\subsection{Class-wise performance analysis}

It is worth mentioning that in total 29 out of 33 cancer types achieved over 90\% accuracy in identification (\tableautorefname~\ref{accuracy per cancer}) with only 500 gene set. 
Specially, BLCA, BRCA, DLBC, GBM, HNSC, KIHC, LAML, MESO, OV, PCPG, PRAD, SKCM, TGCT, THCA, THYM, and UVM, achieved accuracy over 98\% and exhibited distinct characteristics that allowed for their identification from other cancer types. 
The proposed architecture proved successful in handling cancer types with limited patient samples. DLBC attained a 97.78\% accuracy rate despite having a sample size of only 47. KICH, MESO, and UVM achieved accuracy rates of 93.96\%, 98.7\%, and 98.67\% respectively, with sample sizes of 66, 87, and 79. This demonstrates that our proposed architecture is capable of handling class imbalance and performs well even with limited sample sizes. 
However, cancers originating from the same tissue origins posed greater difficulty in distinguishing them from each other compared to those originating from different lineages. For instance, READ and COAD were more challenging to differentiate than BRCA and COAD. Notably, more than half of the READ samples were misclassified as COAD samples, but combining these samples improved the accuracy from 96.39\% to 97.36\% in logistic regression. 

\footnotesize
\begin{longtable}[c]{clc}
\caption{Accuracy of each cancer type of the proposed pipeline.}
\label{accuracy per cancer}\\
\toprule
\textbf{\#} & \textbf{Cancer Type} & \textbf{Acc(\%)}\\ \midrule
\endhead
\midrule
\endfoot
\endlastfoot
1 & Adrenocortical carcinoma (ACC) & 96.00                         \\
2 & Bladder Urothelial Carcinoma (BLCA) & 98.53                        \\
3 & Breast invasive carcinoma (BRCA)  & 99.55                       \\
4 & \begin{tabular}[c]{@{}l@{}}Cervical squamous cell carcinoma\\ endocervical adenocarcinoma (CESC)\end{tabular} & 94.46                        \\
5 & Cholangiocarcinoma (CHOL)  & 81.07                         \\
6 & Colon adenocarcinoma (COAD) & 89.23                        \\
7 & Diffuse large B cell lymphoma (DLBCL)  & 97.78                         \\
8 & Esophageal carcinoma (ESCA)  & 86.28                        \\
9 & Glioblastoma multiforme (GBM) & 98.79 \\
10 & Head and Neck squamous cell (HNSC) & 98.46 \\
11  & Kidney Chromophobe (KICH)  & 93.96 \\
12 & Kidney renal clear cell carcinoma (KIRC)  & 96.42 \\
13  & Kidney renal papillary cell carcinoma (KIRP)  & 93.77 \\
14 & Acute myeloid leukemia (LAML) & 100.00 \\
15 & Brain Lower Grade Glioma (LGG) & 98.66 \\
16 & Liver hepatocellular carcinoma (LIHC)  & 97.04 \\
17 & Lung adenocarcinoma (LUAD) & 95.34 \\
18 & Lung squamous cell carcinoma (LUSC) & 94.97 \\  
19 & Mesothelioma (MESO) & 98.71 \\
20 & Ovarian serous cystadenocarcinoma (OV)  & 99.76 \\
21 & Pancreatic adenocarcinoma (PAAD) & 97.21 \\
22 & Pheochromocytoma and Paraganglioma (PCPG) & 98.90 \\
23 & Prostate adenocarcinoma (PRAD) & 100.00 \\
24 & Rectum adenocarcinoma (READ)  & 42.34 \\
25 & Sarcoma (SARC) & 93.93                        \\
26 & Skin Cutaneous Melanoma (SKCM) & 98.51 \\
27 & Stomach adenocarcinoma (STAD) & 93.49  \\
28 & Testicular Germ Cell Tumors (TGCT)  & 99.26 \\
29 & Thyroid carcinoma (THCA) & 100.00 \\
30 & Thymoma (THYM)  & 98.33 \\
31 & Uterine Corpus Endometrial Carcinoma (UCEC)  & 92.28 \\
32 & Uterine Carcinosarcoma (UCS) & 78.79 \\
 33 & Uveal Melanoma (UVM) & 98.67 \\ \bottomrule
\end{longtable}

\normalsize

In conclusion, the experiments highlighted the significance of appropriate preprocessing approaches in improving model performance while considering computational constraints. The adoption of StandardScaler normalization, SelectFromModel with 500 features, and ensembling of the top three classifiers struck a balance between accuracy and resource efficiency. 


\subsection{Performance Comparison with State-of-the-Art Approaches}

A comparative analysis was conducted with various state-of-the-art \citep{hsu2018cancer,lyu2018deep,de2019deepgx} techniques that utilized the TCGA pan-cancer dataset to perform classification tasks on 33 distinct cancer categories.
\tableautorefname~\ref{sota} displays the performance evaluation for each of the architectures, including the architecture proposed by us. To the best of our knowledge, the proposed architecture outperforms all existing methodologies.

\begin{table}[ht]
\centering
\caption{Performance comparison with state-of-the-art works on 33 types cancer classification using RNA sequence data.}
\label{sota}
\footnotesize
\begin{tabular}{L{7cm} C{3cm} C{1cm}}
\toprule
\textbf{Approach} & \textbf{Number of Genes} & \textbf{Avg. Acc(\%)} \\ \midrule
Linear SVM \cite{hsu2018cancer} & 9900 & 94.98 \\  
Deep learning based algorithm \cite{lyu2018deep} & 10381 & 95.59 \\  
Deep learning based algorithm \cite{de2019deepgx} & 20531 & 95.65 \\ 
\textbf{Ours} & \textbf{500} & \textbf{96.61} \\ \bottomrule
\end{tabular}
\end{table}
\normalsize

One crucial advantage of the proposed pipeline is its ability to achieve superior performance while utilizing a significantly reduced number of features. Our pipeline utilized only 500 features, which is considerably less than other methods. Reducing the number of features greatly reduces the computational resources required for classification tasks. The proposed architecture demonstrates high-performance accuracy and efficient usage of computational resources, making it a promising solution for multi-class cancer classification.

\subsection{Cancer Specific Gene Set}
For the identification of gene biomarkers responsible for specific cancers, we considered the SHAP values of the three best-performing classifiers (XGBoost, Logistic Regression, and SVM).

As discussed previously, we employed a 5-fold cross-validation approach and utilized SelectFromModel to select 500 features in each fold for each classifier. So it is important to note that the 500 selected features in each fold were not exactly identical across all the 5 folds, as they were determined based on the specific data partition in each iteration. Consequently, we obtained SHAP values of features from the 5 folds for each classifier. To obtain a comprehensive global interpretation from each classifier, we combined the SHAP values from the 5 folds. It resulted in a collection of 1008, 1595, and 723 distinct features and their SHAP values for Logistic Regression, XGBoost, and SVM, respectively. 
After the processing of these values as described in  \sectionautorefname~\ref{Global interpretation}, we got 33 distinct sets of feature attribution values specific to the 33 types of cancers from each of the classifiers. We sorted the genes in descending order based on their significance within each cancer type to get the most important genes. 

In order to validate the significance of the genes identified through global interpretation, we conducted a comparison with the differentially expressed genes (DEGs) available for 17 cancer types obtained from Deseq2 analysis. For each cancer, we took the top genes same as the number of the DEGs from our obtained gene sets. From the analysis, we observed that Logistic Regression, XGBoost, and SVM were able to identify a significant number of important genes for most of the cancers. The comparison results are shown in \tableautorefname~\ref{deseq2 intersect}. For instance, in the case of BLCA cancer, the Logistic Regression model identified 97 common genes, XGBoost identified 49 common genes, and SVM identified 122 common genes, all of which were shared with the corresponding DGE. With these findings, we can consider that the genes identified as common with the DEGs are of utmost significance for each cancer type.

We further validated the specificity of the class-specific genes obtained in our study. Typically, if a gene set is truly specific to a particular cancer, it should not exhibit any overlap with genes from a different cancer set. In other words, the number of overlapping genes between different cancer types should be minimal, approaching zero. Our findings indicate that there is indeed a small number of overlapping genes among different cancer types. This outcome serves as validation for the cancer-specific nature of our gene sets.

\begin{table}[t]
\centering
\caption{Number of common gene biomarkers between the gene sets obtained from our pipeline and the Differential Gene Expression from DESeq2.}
\label{deseq2 intersect}
\footnotesize
\begin{tabular}{C{0.5cm} L{1cm} C{3cm} C{3cm} C{3cm}}
\toprule
    \multirow{2}{*}{\#} & 
    \multirow{2}{*}{Cancer} & 
    \multicolumn{3}{c}{\begin{tabular}[C{9cm}]{@{}c@{}}Number of common features between our \\ selected features using SHAP analysis and DEGs \end{tabular}} \\ 
    \cmidrule(l){3-5} 
    & & Logistic Regression    & XGBoost   & Support Vector Machine (SVM) \\ \midrule
1  & BLCA         & 97 & 49 & 122                                                     \\
2  & BRCA         & 60 & 31 & 87                                                     \\
3  & CHOL         & 112 & 233 & 203        \\
4  & COAD         & 112 & 79 & 159        \\
5  & ESCA         & 90 & 57 & 132          \\
6  & HNSC         & 120 & 58 & 133        \\
7  & KICH         & 139 & 159 & 183        \\
8  & KIRC         & 82 & 57 & 108         \\
9  & KIRP         & 74 & 48 & 99         \\
10 & LIHC         & 80 & 43 & 86         \\
11 & LUAD         & 155 & 97 & 200        \\
12 & LUSC         & 239 & 240 & 246        \\
13 & PRAD         & 19 & 7 & 25         \\
14 & READ         & 102 & 76 & 132         \\
15 & STAD         & 71 & 40 & 112        \\
16 & THCA         & 19 & 23 & 52         \\
17 & UCEC         & 145 & 135 & 178        \\ \bottomrule
\end{tabular}
\end{table}

\section{Conclusion}
Gene expression analysis from mRNA data is a valuable tool for cancer classification, leveraging the altered expression levels of specific genes to differentiate between different types of cancer \cite{alba2007gene}. In our study, we introduced a method that efficiently and accurately identifies cancer types and the corresponding gene sets based on mRNA gene expression data. By employing appropriate feature selection techniques, we achieved a significant reduction in computational cost while maintaining a high accuracy rate of 96.61\%. This was accomplished through an ensemble approach incorporating three well-performing classifiers: Logistic regression, Support Vector Machine (SVM), and XGBoost. Furthermore, we computed SHAP scores for each feature to identify and prioritize the most important biomarker genes specific to each cancer type. Comparing these genes with those obtained from DGE analysis underscored the effectiveness of our approach in accurately capturing cancer-specific genes. Overall, our method offers a precise and rapid means of cancer classification while highlighting biologically relevant genes associated with each cancer type.
This study forms the groundwork to identify biomarker genes for individual cancer patients. Analyzing patient-specific gene sets can play an important role in facilitating personalized and targeted treatment approaches.

\bibliographystyle{elsarticle-num} 
 \bibliography{sample}
\end{document}